\documentclass[
 reprint,
 amsmath,
 amssymb,
 aps,
 prl, %
 superscriptaddress
]{revtex4-2}
\usepackage[utf8]{inputenc}
\usepackage{graphicx}
\usepackage{hyperref}
\usepackage{gensymb}
\usepackage{todonotes}
 \setuptodonotes{inline}  

\newcommand{\mrm}{\mathrm}

\newcommand{\Evec}{\vec{\mathcal{E}}}

\newcommand{\Bvec}{\vec{\mathcal{B}}}

\newcommand{\Esca}{\mathcal{E}}

\newcommand{\Bsca}{\mathcal{B}}

\begin{document}

\title{Comagnetometry using mirror-symmetric ions in a crystal}

\author{Bassam Nima}
\author{Mingyu Fan}
\author{Aleksandar Radak}
\affiliation{Department of Physics, University of Toronto, Toronto ON M5S 1A7, Canada}
\author{Andrew Jayich}
\affiliation{Department of Physics, University of California - Santa Barbara, Santa Barbara CA 93106, USA}
\author{Amar Vutha}
\affiliation{Department of Physics, University of Toronto, Toronto ON M5S 1A7, Canada}
\begin{abstract}
Searches for physics beyond the Standard Model using spin sensors are susceptible to spurious frequency shifts and noise due to magnetic fields. Therefore a comagnetometer---an auxiliary sensor that allows mundane magnetic field effects to be differentiated from new physics---is an essential feature of many precision searches. Here we demonstrate the operation of a novel type of comagnetometer using nuclear spins of dopant ions in a crystal, comparing four different sub-ensembles of ions. We demonstrate rejection of magnetic-field-induced shifts to better than 1 part in 10$^5$ using this system, laying the groundwork for improved searches of time-reversal symmetry violation using solid-state systems. 
\end{abstract}

\maketitle

The inconsistencies between the Standard Model of particle physics and astrophysical evidence, such as dark matter and the dominance of matter over antimatter in the universe, motivate searches for new physics that breaks time-reversal symmetry (T) \cite{Cairncross2019}. Precision measurements searching for T-violation in laboratory experiments typically rely on measurements of energy differences between spin states, whether in particles such as neutrons \cite{Abel2020}, atoms \cite{Graner2016} or polar molecules \cite{acme_collaboration_improved_2018,Roussy2023}. However, the spins of these systems are always associated with magnetic moments, which makes such experiments susceptible to magnetic fields: fluctuating magnetic fields add noise and degrade the precision of spin sensors, whereas magnetic fields correlated with changes in experiment parameters can introduce systematic errors. Therefore, a strategy commonly employed in such precision measurements is to use a different spin system as a comagnetometer, which tracks magnetic field changes \cite{Terrano2022}.

Comagnetometer techniques used in searches for new physics range from the use of an entirely different species (e.g., Hg as a comagnetometer in neutron electric dipole moment searches \cite{Abel2020}), to the use of different internal states within the same system (e.g., $\Omega$-doublet states in polar molecule searches for the electron electric dipole moment \cite{DeMille2001,Roussy2023}). Unsurprisingly, any fluctuations in the magnetic moment of the comagnetometer species compared to the system of interest, or variations in their spatial distributions, introduce ways for residual systematic errors to enter into precision measurements \cite{Terrano2022}. An ideal comagnetometer would have an unvarying magnetic moment and be identically distributed in space as the system used for precision measurement, yet have a significantly different sensitivity to new physics effects.

Recently, we initiated work on nuclear T-violation searches using spectroscopy of rare-earth ions in solids, in order to take advantage of the significantly larger number of trapped polarized atoms that are available in doped crystals compared to traditional beams or laser-cooled atomic and molecular experiments. While rare-earth-doped crystals with highly-coherent optical transitions have been employed for quantum information storage \cite{ortu_storage_2022, oswald_characteristics_2018} and laser frequency stabilization \cite{Chen2011,Cook2015}, they can also be used to explore the parameter space of new physics occurring at extreme energy scales ($>$ 100 TeV). In Refs.\ \cite{Ramachandran2022,ramachandran_nuclear_2023}, methods were proposed for performing T-violation searches using non-centrosymmetric crystals containing octupole-enhanced rare-earth nuclei, such as $^{153}$Eu, which are expected to be highly sensitive to new physics. When rare-earth ions are in non-centrosymmetric sites in crystals, they are strongly electrically polarized by the local fields produced by neighboring ions: these polarized ions produce measurable frequency shifts due to T-violating nuclear moments, and they also offer convenient features for mitigating noise and systematic errors. In particular, different groups of ions in these crystals can be used as comagnetometers to control magnetic field effects in T-violation search experiments, as we demonstrate in this Letter.

The essential idea behind comagnetometry in such crystals is to use mirror-symmetric sets of ions that have identical magnetic properties but opposite T-violation sensitivities. In spirit and intent, our work follows the beautiful experiments of Royce and Bloembergen \cite{Royce1963}, who used oppositely-polarized Cr$^{3+}$ ions in crystals to set one of the earliest experimental bounds on T-violation. Here we show that this basic idea can be extended and applied to crystals containing octupole-enhanced nuclei, resulting in comagnetometers with nearly ideal properties.

Our experiments use $^{153}$Eu$^{3+}$ ions doped into yttrium orthosilicate (YSO). In the following, we use $x,y,z$ to refer to the $D1, D2, b$ dielectric axes of the crystal for convenience of notation, and we restrict our attention to ``site 1'' ions in YSO \cite{Yano1991}. Primitive cells in YSO contain four equivalent non-centrosymmetric Y$^{3+}$ ion positions where Eu$^{3+}$ can be substituted. There are two symmetry operations that relate these positions: an inversion,
\begin{equation*}
\hat{\Pi}: (x,y,z)\rightarrow (-x,-y,-z),
\end{equation*} 
and a reflection in the $xy$ plane, 
\begin{equation*}
\hat{\Sigma}: (x,y,z)\rightarrow (-x,-y,z).
\end{equation*}
Consequently, the Hamiltonians for Eu$^{3+}$ ions at these four positions are related to one another. Using the symbols  $\pi = \pm 1$ and $\sigma = \pm 1$ to  label the ion positions according to the irreducible representations of the $\hat{\Pi}$ and $\hat{\Sigma}$ symmetries respectively, the effective Hamiltonians for the nuclear spin degree of freedom at the four positions, in a given electronic state of the ion, are related as  
\begin{equation}\label{eq:hamiltonian}
\begin{split}
    H(\sigma, \pi) =  & \sum_{i,j} Q_{ij} I_i I_j - \sigma \, (\mu_x \Bsca_x + \mu_y \Bsca_y) - \mu_z \Bsca_z \\
    & - \pi D \hat{n} \cdot \vec{\mathcal{E}} \, \mathbb{I} + \sigma \pi \, W \, \vec{I} \cdot \hat{n}.
\end{split}
\end{equation}
The  indices $i,j = 1$ to $3$ label the cartesian axes, and $\Evec$ and $\Bvec$ are respectively the electric and magnetic fields applied to the Eu:YSO crystal. The nuclear spin is $\vec{I}$, the electric quadrupole moment tensor is $Q_{ij}$, and $\mu_i = -\sum_{j} M_{ij} I_j$ are the magnetic moment components in terms of the gyromagnetic tensor $M_{ij}$ \cite{Longdell2006}. We have introduced the quantity $\vec{D} = D \hat{n}$ for the electric dipole moment of the polarized Eu$^{3+}$ ion, with $\mathbb{I}$ being the identity operator for the nuclear spin degree of freedom . The $- \pi D \hat{n} \cdot \vec{\mathcal{E}} \, \mathbb{I}$ term is therefore a uniform energy shift for all the nuclear spin sublevels in an electronic state of the ion, with a sign  that depends on the ion's electric polarization. The quantity $W$ parametrizes the energy shift due to a possible nonzero P-odd, T-odd nuclear moment that is produced by microscopic T-violation. 

From the Hamiltonians in Eq.\ (\ref{eq:hamiltonian}), it is evident that ions related by the $\hat{\Pi}$ transformation have identical quadrupole and Zeeman shifts, but their opposite electrical polarization---due to their mutually inverted local environments---endows them with opposite sensitivity to the $W$-term. By taking advantage of rf-optical double resonance, nuclear spin transitions in ions with different values of $\pi$ can be separately read out using Stark shifts of their optical transitions (due to the $\pi D \hat{n} \cdot \vec{\mathcal{E}} \, \mathbb{I}$ term). In contrast, nuclear spin transitions in ions with different values of $\sigma$ have different resonance frequencies when the magnetic field is applied, e.g., along the $x$ and $z$ directions (due to the $\sigma \mu_x \Bsca_x + \mu_z \Bsca_z$ term), but they have identical shifts due to electric fields. As we show below, these two types of behavior can be used to isolate the $\sigma \pi$-dependent $W$-term,  which contains information about possible beyond-Standard-Model physics, and separate it from spurious magnetic field effects.

Crucially, all these measurements can be made just by changing optical and rf frequencies, without requiring mechanical or electrical switches that could introduce correlated systematic errors. The $\sigma=\pm1, \pi=\pm1$ sub-ensembles of Eu$^{3+}$ ions therefore act as effective comagnetometers for each other. In the following, we show that such a scheme can be experimentally implemented and practically used in high-precision measurements.

\begin{figure}[h!]
    \centering
    \includegraphics[width=\linewidth]{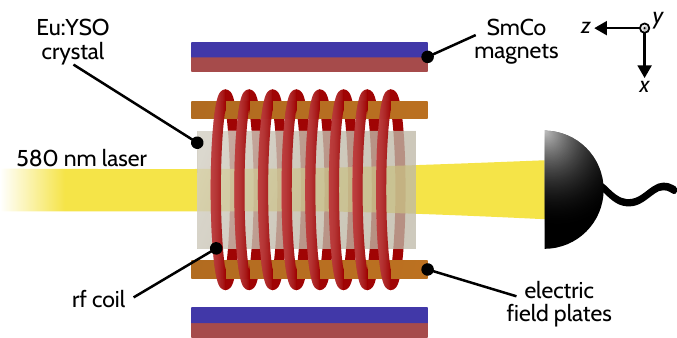}
    \caption{Schematic diagram of the apparatus. The experimental data shown in Figs.\ \ref{fig:ramsey} and \ref{fig:comagnetometry_data} were taken using laser absorption spectroscopy of the $^7F_0 - {}^5D_0$ transition in a Eu:YSO crystal, held at 5 K. The crystal was placed between a pair of electric field plates, contained within an rf magnetic field coil. A static magnetic field parallel to the electric field was applied using SmCo permanent magnets.}
    \label{fig:setup}
\end{figure}

\begin{figure}[h!]
    \centering
    \includegraphics[width=1\columnwidth]{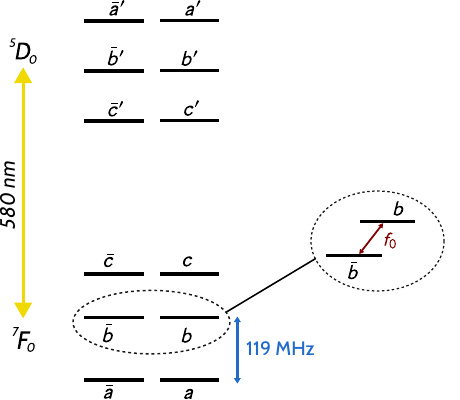}
    \caption{Energy levels of $^{153}$Eu$^{3+}$:YSO. The ground ${}^7F_0$ and excited ${}^5D_0$ electronic states are connected by a 580 nm optical transition. In measurements reported here, the $b-\bar{b}$ nuclear spin transition was probed using rf spectroscopy. The energy difference $f_0$ depends mainly on the applied magnetic field, along with a small contribution due to possible T-violation in the $^{153}$Eu nucleus. 
    }
    \label{fig:structure}
\end{figure}

The energy level structure of Eu:YSO is shown in Fig.\ \ref{fig:structure}. The 580 nm optical transition connects the ground $^7F_0$ electronic state to the excited $^5D_0$ electronic state in Eu$^{3+}$. The quadrupole interaction splits the 6 nuclear spin sublevels into 3 sets of Kramers-conjugate state pairs \cite{Longdell2006}, which we label as $\bar{a},a; \bar{b},b; \bar{c},c$ in the $^7F_0$ electronic state, and $\bar{a}',a'; \bar{b}',b'; \bar{c}',c'$ in the $^5D_0$ electronic state.

The T-violation sensitivities of the $\bar{a},a; \bar{b},b; \bar{c},c$ states are discussed in Ref.\ \cite{ramachandran_nuclear_2023}. For the measurements reported here, we focus on the $b \to \bar{b}$ transition, which can be used to search for nuclear T-violation \cite{ramachandran_nuclear_2023}. The resonance frequency of this transition has a magnetic field sensitivity $\partial f(b,\bar{b})/ \partial \mathcal{B}_x \approx$ 0.6 kHz/G. Therefore, without comagnetometry and using magnetic field control alone, it is challenging to measure resonance frequency shifts due to beyond-Standard-Model T-violation that are expected to be $\lesssim$ 1 mHz. 

\begin{figure}[h!]
    \centering
    \includegraphics[width=\columnwidth]{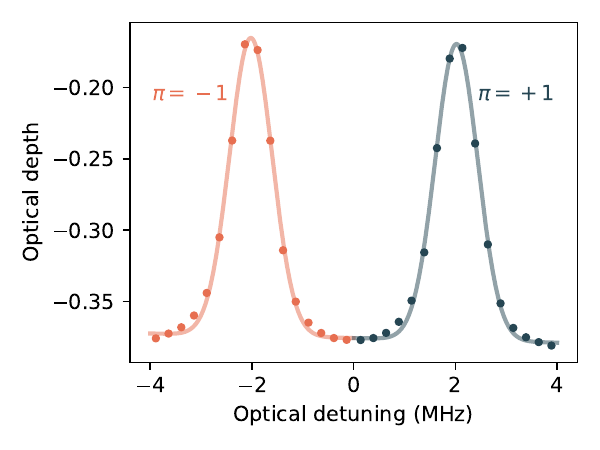}
    \caption{Optical spectrum with $\Esca_\mrm{dc} \approx 70$ V/cm, used to resolve the $\pi=\pm 1$ sub-ensembles. The optical depth is proportional to the population in the $a, \bar{a}$ hyperfine states.}
    \label{fig:optical}
\end{figure}

\begin{figure}[h!]
    \centering
    \includegraphics[width=1.02\columnwidth]{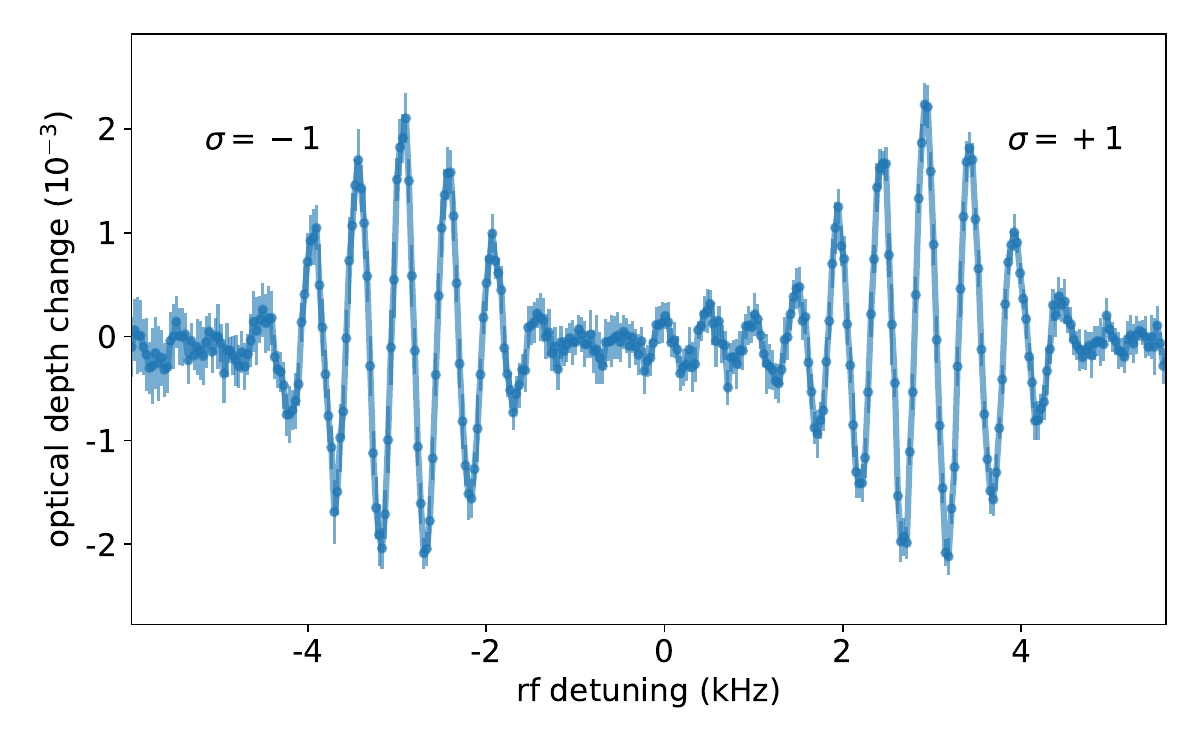}
    \caption{Ramsey spectroscopy of the $b-\bar{b}$ transition for the $\sigma = \pm 1, \pi = +1$ sub-ensembles. The solid line through the data points is a guide to the eye. Eu$^{3+}$ ions were initially prepared in the $b$ state using optical pumping, and the transition probability measured using laser absorption as the carrier frequency of the Ramsey pulses is scanned. The rf detuning is relative to 223.0 kHz.} 
    \label{fig:ramsey}
\end{figure}

The YSO crystal in our experiments is doped with $^{153}$Eu$^{3+}$ at a concentration of 0.01\%, and has dimensions 3.5 mm 
$\times$ 4.0 mm 
$\times$ 5.0 mm. The crystal was mounted in a cryocooler and held at approximately 5 K. A pair of silver coated plates adjacent to the crystal were used to apply an electric field $\Esca_\mrm{dc}=70$ V/cm along the $\hat{x}$ axis, which shifted the optical absorption resonances from the $\pi = \pm1$ sub-ensembles by $\Delta \nu_\Esca = \pm 2$ MHz, as shown in Fig.\ \ref{fig:optical}.  A solenoid around the crystal and field plates was used to generate high-frequency rf magnetic fields at 119 MHz along the $\hat{z}$ axis 
in order to drive the $a,\bar{a} \to b,\bar{b}$ transition in the $^7F_0$ ground electronic state, for state preparation and detection. A pair of coils aligned along the $\hat{x}$ axis were used to apply low-frequency ($\sim$200 kHz) rf fields to drive the $b \to \bar{b}$ transition for spectroscopy. Samarium-cobalt permanent magnets were used to apply a static magnetic field, $\Bsca_x \approx 350$ G, across the Eu:YSO crystal: we chose this value of $\Bsca_x$ in order to resolve the $a \to b$ and $\bar{a} \to \bar{b}$ resonances on the 119 MHz transition. External coils were used to apply a small static magnetic field $\Bsca_z \approx$ 7 G to resolve the $\sigma = \pm 1$ sub-ensembles, shifting their resonances by $\Delta f_\mathcal{B} = \pm 3$ kHz, as shown in Fig.\ \ref{fig:ramsey}.

\begin{figure*}
    \centering
    \includegraphics[width=2.06\columnwidth]{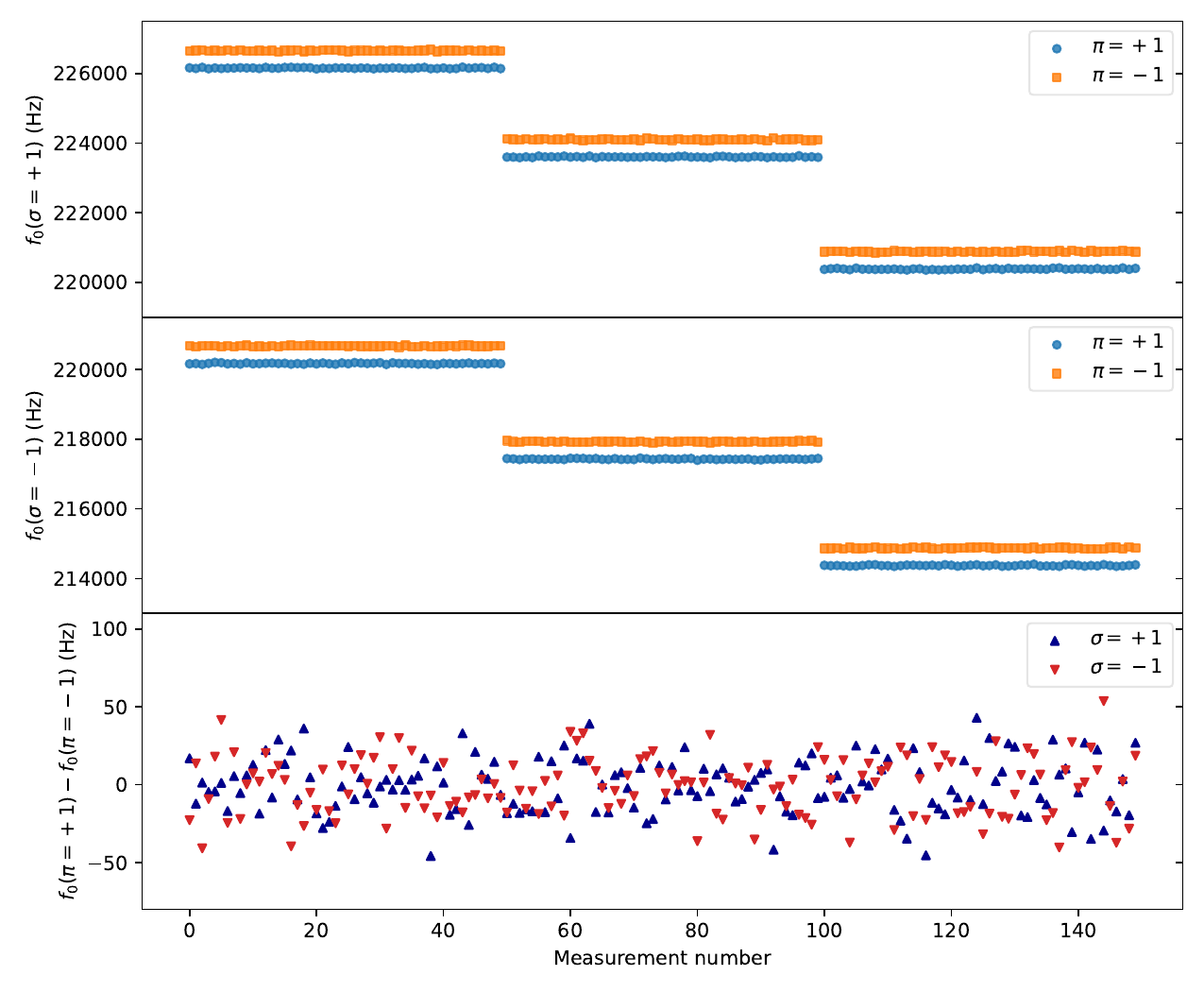}
    \caption{Resonance frequencies $f_0(\sigma,\pi)$ for the $b-\bar{b}$ transition in the $\sigma=\pm1, \pi=\pm1$ sub-ensembles. In the top two panels, the data points  for the $\pi = -1$ sub-ensemble are shown shifted by an artificial offset of +500 Hz, as otherwise they are imperceptibly close to the $\pi=+1$ resonance frequencies. Magnetic field steps were applied three times during this dataset leading to the frequency jumps -- despite these large perturbations the $\pi=\pm1$ resonance frequencies accurately track each other. The bottom panel zooms in on the differences between $\pi=\pm1$ resonance frequencies for $\sigma=+1$ and $\sigma=-1$ sub-ensembles, demonstrating the excellent rejection of common-mode frequency shifts from magnetic fields.}
    \label{fig:comagnetometry_data}

\end{figure*} 

With the $\Esca$-field on, we used optical and rf pulses to initialize the $^{153}$Eu ions into a single nuclear spin state. A sequence of laser pulses were used to drive the ${}^7F_0 \ a, \bar{a} \to {}^5D_0 \ c', \bar{c}'$ and ${}^7F_0 \ c, \bar{c} \to {}^5D_0  \ b',\bar{b}'$ transitions, while  rf pulses drove the 119 MHz $\bar{b} \to \bar{a}$ transition within the ground ${}^7F_0$ electronic state. The state-preparation sequence follows the method used in Ref.\ \cite{cruzeiro_characterization_2018}, with the only difference being a final $\bar{b} \to \bar{a}$ rf pulse to prepare ions in the $b$ state. 

Following  state preparation, the $\Esca$-field was switched off and rf spectroscopy of the $b \to \bar{b}$ transition was performed using the Ramsey method \cite{Ramsey1950}. Two pulses of duration $\tau =$ 20 $\mu$s separated by $T =$ 80 $\mu$s were applied to the crystal, and the carrier frequency of these pulses was varied. The population excited by these pulses to the $\bar{b}$ state was subsequently transferred to the $a$ state using an adiabatic sweep pulse, after which the population in the $a$ state was measured using optical absorption on the ${}^7F_0 \ a,\bar{a} \to {}^5D_0 \ c',\bar{c}'$ transition. Since the state-preparation step is performed in the presence of an $\Esca$-field, two resolved optical absorption lines corresponding to the $\pi = \pm1$ sub-ensembles can be observed on the ${}^7F_0 \ a,\bar{a} \to {}^5D_0 \ c',\bar{c}'$ transition, as shown in Fig.\ \ref{fig:optical}. Measurements of the optical depth of these absorption lines, as the carrier frequency of the spectroscopy pulses was scanned across the $b \to \bar{b}$ transition, yielded two resonance frequencies $f_0(\sigma = \pm1,\pi)$ for each value of $\pi$, as shown in Fig.\ \ref{fig:ramsey}.

The four resonance frequencies were repeatedly measured as shown in Fig.\ \ref{fig:comagnetometry_data}. The line centers vary over time due to background magnetic field drifts of around 0.1 G in our unshielded laboratory. We also intentionally applied relatively large $\pm3$ G step perturbations to the magnetic field along the $x$-axis, in order to amplify any systematic errors produced by magnetic fields. Crucially, the resonances for the $\sigma=\pm 1, \pi = \pm1$ sub-ensembles precisely track each other, as shown in Fig. \ref{fig:comagnetometry_data}.  As can be seen in the figure, for each value of $\sigma$ the resonances for the $\pi = \pm 1$ sub-ensembles are identical to within the uncertainty of our measurements. We can quantify the deviations in two ways using the $\sigma=\pm1$ sub-ensembles: for each value of $\sigma$, we compute the fractional difference $y(\sigma) = [f_0(\sigma,\pi=+1) - f_0(\sigma,\pi=-1)] / f_\mrm{avg}$. We find $y(\sigma=+1) = (-5 \pm 7) \times 10^{-6}$ and $y(\sigma=-1) = (-4 \pm 5) \times 10^{-6}$, both of which are consistent with zero. We thus confirm that oppositely-polarized Eu$^{3+}$ ions in Eu:YSO have identical magnetic moments and can cancel magnetic field effects to better than 10 parts-per-million: mirror-symmetric ions in the crystal act as excellent mutual comagnetometers. 

Importantly, whereas the comagnetometer ions have correlated magnetic field shifts, they have \emph{opposite} sensitivities to T-violating new physics, as can be seen from the $\sigma \pi$ dependence of the $W$ term in Eq.\ \ref{eq:hamiltonian}.  Thus errors and noise due to magnetic fields, which lead to common-mode frequency shifts of the comagnetometer ions, can be cleanly distinguished from the differential shifts produced by T-violating new physics. We  anticipate that the methods demonstrated in this paper can be used to significantly improve the precision and accuracy of nuclear T-violation measurements using solid-state systems.

In summary, we have performed precision comagnetometry using the nuclear spins of mirror-symmetric ions in a crystal, demonstrating that shifts in their nuclear spin transitions due to magnetic field errors can be accurately canceled. Our work shows that ions trapped in crystals have useful properties to advance the precision and accuracy of searches for T-violation beyond the Standard Model. 

~\\
\emph{Acknowledgments.} We thank Yoshiro Takahashi, Jonathan Weinstein, Eric Hessels and Philippe Goldner for helpful discussions. Julia Ford and Daniel Stedman contributed to the construction of the initial apparatus, and Harish Ramachandran made significant contributions to developing the T-violation search concept. We are grateful to Bob Amos and Paul Woitalla for vital technical support. M.F. acknowledges funding from a CQIQC Postdoctoral Fellowship, and A.R. acknowledges funding from a CQIQC Undergraduate Summer Research Award. This project was enabled by support from the John Templeton Foundation (Grant No. 63119), the Alfred P. Sloan Foundation (Grant No. G-2023-21045), and NSERC (SAPIN-2021-00025).

\bibliography{eu_comag}
\end{document}